\documentclass[a4,12pt]{article}
\pdfoutput=1
\usepackage{amsmath,amssymb,color,graphics,amscd,amsfonts,epsf,indentfirst}
\allowdisplaybreaks[4]

\makeatletter
\@addtoreset{equation}{section}

\makeatletter
\renewcommand\section{\@startsection {section}{1}{\z@}%
                                   {-3.5ex \@plus -1ex \@minus -.2ex}
                                  {2.3ex \@plus.2ex}%
                                   {\normalfont\large\bfseries}}
\renewcommand\subsection{\@startsection{subsection}{2}{\z@}%
                                     {-3.25ex\@plus -1ex \@minus -.2ex}%
                                     {1.5ex \@plus .2ex}%
                                    {\normalfont\bfseries}}

\def\btab{\begin{table}[h] \begin{center} \begin{tabular}{l lp{3in}}}
      \def\etab{\end{tabular} \end{center} \end{table}}
\def\btabm{\begin{center} \begin{tabular}}
    \def\etabm{\end{tabular} \end{center}}

\newcommand{\be}{\begin{equation}}
\newcommand{\ba}{\begin{eqnarray}}
\newcommand{\ea}{\end{eqnarray}}
\newcommand{\ee}{\end{equation}}

\begin{document}

\begin{titlepage}
  \thispagestyle{empty}

\begin{flushright}
\rightline{WITS-CTP-098}
  \end{flushright}
  \vspace{1cm}
  \begin{center}
   \font\titlerm=cmr10 scaled\magstep4
   \font\titlei=cmmi10 scaled\magstep4
   \font\titleis=cmmi7 scaled\magstep4

\centerline{\titlerm Lightlike contours with fermions
\vspace{1.5cm}}
\noindent{{
       Pawe{\l} Caputa$\,\,$\,
       }}\\
    \vspace{1.0cm}

    {\it  National Institute for Theoretical Physics,\\
 Department of Physics and Centre for Theoretical Physics,\\
 University of the Witwatersrand, Wits, 2050,\\ South Africa}
\vspace{0.7cm}

    {\it \it \text{\rm pawel.caputa@wits.ac.za}}

  \end{center}

\vskip 5em

  \begin{abstract}
We consider high energy scattering of open superstrings in flat spacetime. It is shown that in this regime amplitudes are dominated by minimal supersurfaces spanned by lightlike supercontours built of the on-shell momenta.  We demonstrate how using a generalization of the functional over reparametrizations of the boundary one can extract the area of the minimal supersurfaces directly from the lightlike contour. Exponent of the area correctly reproduces RNS amplitudes. 
We also comment on possible implementation of the supercontours to Wilson loops/Scattering amplitudes duality.
  \end{abstract}

\end{titlepage}




\section{Introduction}

Scattering amplitudes in string theory are computed by path integrals weighted by the exponent of minus the area of the string's worldsheet. Therefore the saddle point approximation to the amplitude is simply given by the area of the minimal surface. This fact was first used by Gross and collaborators \cite{Gross:1987kza,Gross:1989ge} to study flat space, closed and open string scattering at high center of mass energy and fixed angle. They showed that precisely in such a regime the relevant information about the process is encoded in a classical minimal surface with spikes that correspond to insertions of appropriate vertex operators.

More recently, Alday and Maldacena \cite{Alday:2007hr} also implemented the connection between string amplitudes and minimal surfaces in Anti-de Sitter space in order to model scattering of strongly coupled gluons in dual $\mathcal{N}=4$ supersymmetric Yang-Mills theory (for more details and further references see \cite{Alday:2008yw}). The key step in their analysis was  a ``trick" of performing T-duality along the four Minkowski directions\footnote{Supplemented by inversion of the radial coordinate $z\to1/z$ }. This operation maps the insertion points of the vertex operators into lightlike segments of on-shell momenta and intervals between them into cusps\footnote{As shown later in \cite{Berkovits:2008ic} such transformations combined with fermionic T-duality leave the full $AdS_{5}\times S^{5}$ background invariant.}. This way, minimal surfaces relevant for gluon scattering at strong coupling can be viewed as soap bubbles in $AdS$ spanned by closed lightlike loops built of the intervals of the physical momenta.

These boundary loops, turned out to have a very interesting implications on the relation between amplitudes and Wilson loops in $\mathcal{N}=4$ SYM. According to the AdS/CFT dictionary, an expectation value of a Wilson loop along contour $C$ corresponds to a minimal surface in AdS that is attached to $C$ at the boundary \cite{Maldacena:1998im,Rey:1998ik}. This way, the result of Alday and Maldacena was first interpreted as a strong coupling duality between Wilson loops and scattering amplitudes (WL/SA duality). Surprisingly this relation was later confirmed for 4 and 5 gluon MHV amplitudes and Wilson loops in perturbation theory \cite{Drummond:2007cf,Brandhuber:2007yx}. Its verification for more than 6 particles, higher loops and beyond MHV amplitudes is still a very active area of research (see e.g. \cite{Bern:2008ap}, \cite{Mason:2010yk,CaronHuot:2010ek,Beisert:2012gb} and \cite{Alday:2010kn} for more details). 

Even though chronologically cusped minimal surfaces first appeared in the AdS/CFT correspondence, by now it is clear that every open string amplitude (disc) at high energies can be obtained from (possibly a generalization) of a lightlike momentum contour. In particular in \cite{Caputa:2011zk}, we investigated the T-duality trick in the simplest possible setup, the bosonic strings in flat space. We found the explicit form of the T-dual worldsheet and verified that it is precisely bounded by a closed contour built of the null on-shell momenta of the scattered states; the same as the one found by Alday and Maldacena. Furthermore we demonstrated that the exponent of minus the area of the cusped lightlike surfaces correctly reproduces the well known open string amplitudes (Veneziano and Bardakci-Ruegg). 

Our main goal in this Letter is to shed some light on how fermions enter to the geometry of the lightlike contours. We believe, that this analysis is closely related to the recent progress in generalizing Wilson loops such that they capture the information about $N^kMHV$ amplitudes. All three approaches to this problem \cite{Mason:2010yk,CaronHuot:2010ek,Beisert:2012gb} seem to suggest that such loops must be supersymmetric. Here we work within the flat space toy model for the Alday-Maldacena construction and we generalize our analysis from \cite{Caputa:2011zk} to superstrings in RNS formulation. More precisely, working in the superfield formulation, we  construct the explicit form of a T-dual superworldsheet and focus on the contour at its boundary. Similarly to the bosonic case it is a lightlike momentum loop, but now in superspace. The loop turns out to be expressed in terms of a supersymmetric step function and it is invariant under local supersymmetry transformations. Curiously, an identical geometrical object was found before by Migdal \cite{Migdal:1996rn} in the context of super-Wilson loops in momentum space. We comment on this relation in the main text. 

This Letter is organized as follows. In the next section we briefly review minimal surfaces that dominate open bosonic string amplitudes at high energies. We sketch how the T-duality transforms the worldsheet and how the lightlike momentum contour emerges at the boundary. In section \ref{SSt} we consider high energy scattering of superstrings approximated by spiky worldsheets in superspace. We explicitly find the T-dual superworldsheet and focus on its boundary. Then we generalize Douglas functional over reparametrizations of the boundary into supersurfaces and apply it to demonstrate that exponent of the area of the surface spanned by our supercontour correctly reproduces the RNS amplitudes. In section \ref{FWL} we comment on possible implications of our analysis for WL/SA duality and finally we end with conclusions. In Appendix \ref{aA} we include a lightning review of the Douglas functional approach to minimal surfaces and derivations of the relevant formulas.

\section{Bosonic amplitudes and T-duality}

In this section, we briefly review how T-duality transforms classical bosonic worldsheets (surfaces) in flat space \cite{Caputa:2011zk}. The key object in our discussion will be the boundary contour that arises after the transformation. Clearly this analysis is motivated by the trick of Alday and Maldacena \cite{Alday:2007hr} and can be thought of as a flat space toy model for their very non-trivial results in $AdS$. Indeed, as we will see below, from the geometrical point of view, the four-dimensional T-dual contour is universal.

Classical worldsheet that describes scattering of $n$ open string tachyons\footnote{At high energies we assume that strings are long and their oscillations are negligible. Therefore scattering amplitudes are insensitive to the asymptotic states and the intercept can be dropped. Strictly speaking this assumption is valid in the Regge regime.} with momenta $k_{i}$ is given by \cite{Polchinski}
\begin{equation}
X^{\mu}(\tau,\sigma) =2i\alpha'\sum^{n}_{i=1}k^{\mu}_{i}\log\sqrt{(\sigma-\sigma_{i})^{2}+\tau^{2}}.\label{Ntachyons}
\end{equation}
It is parametrized on the upper half plane: $\sigma\in(-\infty,\infty)$, $\tau\geq 0$, and boundary points $(0,\sigma_{i})$ correspond to insertions of the vertex operators. The worldsheet represents a surface with n spikes where $X^{\mu}(0,\sigma_{i})\sim k^{\mu}_{i}\log(\sigma-\sigma_{i})$. We require that momenta in the scattering process are conserved and null, so we have 
\begin{equation}
\sum^{n}_{i=1}k^{\mu}_{i}=0,\qquad k^{2}_{i}=0.\label{conservation}
\end{equation}

Let us now analyze how T-duality transforms \eqref{Ntachyons}. Fortunately, in flat space, this can be done in great detail since it is straightforward to construct the dual worldsheet $Y^{\mu}(\tau,\sigma)$. This is done by solving the equations obtained from the standard Buscher procedure \cite{Buscher:1987qj}
\begin{eqnarray}
\partial_{\tau}Y^{\mu}&=&\partial_{\sigma}X^{\mu}\nonumber\\
\partial_{\sigma}Y^{\mu}&=&-\partial_{\tau}X^{\mu},
\label{TDUALITY}
\end{eqnarray}
which one can view as Cauchy-Riemann equations for some analytic function $Z^{\mu}=Y^{\mu}+i\,X^{\mu}$. In our case it turns out to be the logarithm so we just use the identity
\begin{equation}
\log(\sigma-\sigma_{i}+i\,\tau)=\log\sqrt{(\sigma-\sigma_{i})^{2}+\tau^{2}}-i\,\arctan\left(\frac{\sigma-\sigma_{i}}{\tau}\right),
\end{equation}
to immediately write the T-dual worldsheet \cite{Caputa:2011zk}
\begin{equation}
Y^{\mu}(\tau,\sigma)=-2i\alpha'\sum^{n}_{i=1}k^{\mu}_{i}\arctan\left(\frac{\sigma-\sigma_{i}}{\tau}\right).
\label{Tdual}
\end{equation}

The main object of our interest is now the contour that spans \eqref{Tdual}, and we read it off after taking the $\tau\to 0$ limit\footnote{The boundary is located at $\tau=0$}. It is given by
\begin{equation}
Y^{\mu}(0,\sigma)=-2\pi i\alpha'\sum^{n}_{i=1}k^{\mu}_{i}\,\theta(\sigma-\sigma_{i}),
\label{contour}
\end{equation}
where $\theta(\sigma)$ is the step function\footnote{We used the identity $\theta(x)=\lim_{\tau\to0}\left(\frac{1}{2}+\frac{1}{\pi}\arctan\frac{x}{\tau}\right)$, and momentum conservation \eqref{conservation}.}.\\
Clearly, as curve \eqref{contour} tells us, T-duality maps insertion points $\sigma_{i}$ of the vertex operators into lightlike segments and distances between them into cusps, such that the difference between two neighboring cusps is proportional to the physical momentum 
\begin{equation}
Y^{\mu}(0,\sigma_{i})-Y^{\mu}(0,\sigma_{i-1})=-2\pi i\alpha' k^{\mu}_{i}.\label{cusps}
\end{equation}
Precisely the same contour was found by Alday and Maldacena \cite{Alday:2007hr} at the boundary of the worldsheet in $AdS_{3}$ and later it was used to compute the expectation values of Wilson loops dual to scattering amplitudes in $\mathcal{N}=4$ SYM \cite{Drummond:2007cf}. 

To complete the story, one can show that area of the minimal surface spanned by \eqref{contour} captures the leading high energy behavior of the open bosonic string amplitudes. There are several ways to compute the area.  A particularly elegant one, uses the functional over reparametrizations of the boundary developed by Douglas (for more details see Appendix \ref{aA}). For surfaces parametrized on the upper half plane the functional is given by
\begin{equation}
D[C_{n}]=\int^{+\infty}_{-\infty}df_{1}\int^{+\infty}_{-\infty}df_{2}\,C'^{\mu}(f_{1})C'_{\mu}(f_{2})\log(\sigma(f_1)-\sigma(f_2)),
\end{equation}
where $f$ is the reparametrization of the boundary and $\sigma$ is its inverse. As shown in \cite{Caputa:2011zk}, if evaluated on the lightlike contour \eqref{contour} it yields
\begin{equation}
\frac{D[C_{n}]}{4\pi^{2}\alpha'}=\alpha'\sum^{n}_{i\neq j}k_{i}\cdot k_{j}\log(\sigma_{i}-\sigma_{j}).\label{bampl}
\end{equation}
hence its exponent correctly reproduces open string amplitudes.\\
To make the connection with \cite{Gross:1987kza} we can further minimize \eqref{bampl} with respect to the insertion points $\sigma_{i}$\footnote{Three of the 
$\sigma$'s can be fixed due to the $Sl(2,\mathbb{R})$ invariance. It is customary to fix $(\sigma_{1},\sigma_{n-1},\sigma_{n})=(0,1,\infty)$} (this is equivalent to imposing Douglas minimization conditions; see Appendix \ref{aA}). For example for four tachyons we only minimize with respect to $\sigma_2$ and the saddle point approximation reproduces the well known soft behavior 
\begin{equation}
A_{4}\sim e^{-\alpha'(s\log s+t\log t+u\log u)},
\end{equation}
where $(s,t,u)$ are the standard Mandelstam variables.

\section{Superstrings at high energies and lightlike contours with fermions} \label{SSt}
In this section we generalize the above discussion to high energy and fixed angle scattering of RNS superstrings. We will package bosonic and fermionic coordinates into one superfield, so by term ``supersurfaces" we will mean classical solutions of the equations of motion that come from the supersigma model.  Our main objective is to construct a T-dual supersurface and determine its boundary contour. This should be a lightlike loop that captures the relevant information about both, bosonic and fermionic parts of the superamplitude. To justify our procedure we will demonstrate that exponent of the area of the cusped superworldsheet reproduces well known RNS amplitudes.

Superstrings in RNS formulation can be elegantly described in terms of a superfield \cite{Friedan:1985ey}
\begin{equation}
X^{\mu}_{s}(z,\bar{z},\theta,\bar{\theta})=X^{\mu}(z)+X^{\mu}(\bar{z})+\theta\, \psi^{\mu}(z)+\bar{\theta}\,\bar{\psi}^{\mu}(\bar{z})+\theta\bar{\theta}\,F^{\mu},
\end{equation}
where $X^{\mu}$ and $\psi^{\mu}$ are the bosonic and the fermionic coordinates respectively and $F^{\mu}$ is an auxiliary field that makes supersymmetry manifest\footnote{For our purposes we can just integrate it out using its equation of motion $F^{\mu}=0$}, and we introduced additional complex Grassmann coordinates $(\theta,\bar{\theta})$.\\
For open superstrings we will focus on z-dependent fields $(X^{\mu}(z),\psi^{\mu}(z))$. The worldsheet will be also parametrized on the upper half plane $z=\sigma+i\tau$ where $\sigma\in(-\infty,+\infty)$, $\tau\in(0,+\infty)$. 

The interactions are governed by the supersigma model\footnote{For simplicity we drop factors of $\pi$ and $\alpha'$}
\begin{equation}
S=\frac{1}{2}\int d^{2}z \,d^{2}\theta\,\bar{D}X^{\mu}_{s}DX^{\nu}_{s}\eta_{\mu\nu},
\label{superA}
\end{equation} 
where superderivatives are defined as
\begin{equation}
D=\partial_{\theta}+\theta\,\partial_{z},\qquad \bar{D}=\partial_{\bar{\theta}}+\bar{\theta}\,\partial_{\bar{z}}.
\end{equation}
To make our point, it is enough to consider the simplest tree-level amplitude for scattering of n superstrings. Therefore our vertex operators are just supersymmetric generalizations of the tachyonic plane-waves \cite{GS}
\begin{equation}
V_{i}(\sigma_{i},\theta_{i})=e^{i\,k_{\mu}\cdot X^{\mu}_{s}(\sigma_{i},\theta_{i})}.
\end{equation}
A disc superamplitude is then computed by
\begin{equation}
A_{n}=\int\prod^{n-2}_{i=2} d\sigma_{i}\int\prod^{n}_{j=1}d\theta_{j}\,\langle\prod^{n}_{l=1}V_{l}(\sigma_{l},\theta_{l})\rangle,
\end{equation}
where the expectation value is taken with action \eqref{superA}.

Similarly to the bosonic case, at high energies the amplitude is dominated by the saddle point that satisfies the following equations of motion
\begin{equation}
\bar{D}D\,X^{\mu}_{s}=i\sum^{n}_{i=1}k^{\mu}_{i}\delta(S-S_{i}),
\label{saddle}
\end{equation}
where terms on the right hand side come from vertex operators that serve as sources at the superboundary. Moreover, we introduced the following notation: $S=(\sigma,\theta)$ represents a boundary point in superspace, a generalized distance between two points at the superboundary is defined to be
\begin{equation}
S-S_i=\sigma-\sigma_i-\theta\,\theta_i,
\end{equation}
and $\delta(S-S_i)$ is a supersymmetric delta function defined by
\begin{equation}
\int dS\, \delta(S-S_{i})\,f(S)=f(S_{i}),
\end{equation}
where we formally denote $dS\equiv d\theta\,d\sigma$.

The solution to \eqref{saddle} is given by \cite{Polchinski}
\begin{equation}
X^{\mu}_{s}(\sigma,\tau,\theta)=i\sum^{n}_{i=1}k^{\mu}_{i}\log\sqrt{(\sigma-\sigma_{i}-\theta\theta_i)^{2}+\tau^{2}}.\label{SolPolch}
\end{equation}
It represents a supersurface in with $n$ spikes corresponding to the insertion points $(\sigma_{i},\theta_{i})$ where the worldsheet scales as
\begin{equation}
X^{\mu}_{s}\sim i\, k^{\mu}_{i}\,\log(\sigma-\sigma_{i}-\theta\theta_{i}).
\end{equation}

\section{T-dual superworldsheet}\label{TdualWS}

In this section we find an explicit parametrization of the T-dual worldsheet. Fortunately in can be obtained by a simple modification of the bosonic counterpart \eqref{Tdual}.  Namely, we just substitute $\sigma-\sigma_{i}\to\sigma-\sigma_{i}-\theta\,\theta_{i}$ in \eqref{Tdual} what brings us to
\begin{equation}
Y^{\mu}_{s}(\sigma,\tau,\theta)=-i\sum^{n}_{i=1}k^{\mu}_{i}\arctan\left(\frac{\sigma-\sigma_{i}-\theta\theta_{i}}{\tau}\right).\label{SuperTdual}
\end{equation}
There are several ways to check that this solution is the super-T-dual of \eqref{SolPolch}. For example, one can just apply the formal Buscher procedure to the supersigma model \eqref{superA} and derive the super-T-duality constraints\footnote{This procedure was applied before to the GS string sigma model in $AdS$ by Berkovits and Maldacena \cite{Berkovits:2008ic}}. We checked that they imply that both, bosonic and fermionic parts separately satisfy the T-duality constraints \eqref{TDUALITY} \footnote{In the case of our solution this can be easily verified by noticing that the two functions that we used above, $\log$ and $\arctan$, are just the real and imaginary parts of $\log(z-z_{i}-\theta\theta_{i})$, where $z=\sigma+i\,\tau$ and $z_{i}=\sigma_{i}$. This way the T-duality equations are just the Cauchy-Riemann equations for real and imaginary parts of $\log(z)$ and $1/z$.}. Equivalently, one can just take the minimal surface \eqref{SuperTdual} and verify that, in a complete analogy with the bosonic case, it leads to the correct form of the superamplitude.

Let us now proceed with analysis of the boundary. Taking $\tau\to 0$ we arrive at
\begin{equation}
Y^{\mu}_{s}(0,S=(\sigma,\theta)) = -\pi i\sum^{n}_{i=1}k^{\mu}_{i}\,\Theta(S-S_{i}),
\label{Scontour}
\end{equation}
where $\Theta$ is a supersymmetric generalization of the step function\footnote{To avid the confusion between the step function and Grassmann variables we denote the former by $\tilde{\theta}$ }
\begin{equation}
\Theta(S-S_{i})\equiv\tilde{\theta}(\sigma-\sigma_{i})-\theta\theta_{i}\delta(\sigma-\sigma_{i}),\label{Stheta}
\end{equation}
that was first introduced in \cite{Andreev:1988cb}. By definition it satisfies
\begin{equation}
D_{i}\,\Theta(S_{i}-S_{j})=\delta(S_{i}-S_{j}),
\end{equation}
where we labeled superderivatives just to distinguish parameters that they differentiate with respect to. Also in everything what follows superderivatives will act on contours that depend on $(\sigma,\theta)$ only, so for all our purposes: $D_{i}=\partial_{\theta_{i}}+\theta_{i}\partial_{\sigma_{i}}$.

It is clear that \eqref{Scontour} is a natural generalization of \eqref{contour}\footnote{It was also pointed to us \cite{Shinji} that the supersymmetric worldsheet and the contour with \eqref{Stheta} can be easily modified to include the information about polarizations of the scattered states. Namely if we replaced $k^{\mu}_{i}\theta\,\theta_{i}\to\epsilon_{i}\,\zeta^{\mu}_{i}$ and differentiated the answer w.r.t $\epsilon_{i}$ we would obtain the saddle point approximation to the (gluon) amplitude with polarizations.}. Now the insertion points $(\sigma_{i},\theta_{i})$ are mapped into segments proportional to lightlike, on-shell momenta and intervals between them into cusps. We then have the analog of \eqref{cusps} given by
\begin{equation}
Y^{\mu}_{s}(0,\sigma_{i},\theta_i)-Y^{\mu}_{s}(0,\sigma_{i-1},\theta_{i-1})=-\pi i\,k^{\mu}_{i},\qquad k^{2}_{i}=0.
\end{equation}

At this point it is also interesting to note that contour \eqref{Scontour} appeared before in the work of Migdal on super-Wilson loops in momentum space and QCD string \cite{Migdal:1996rn}. There \eqref{Scontour} was defined as a ``supermomentum" and had a direct link to the correlation functions of external flavor currents.\\ 
In addition Migdal showed that such a loop is invariant under local SUSY
\begin{equation}
\delta\sigma=\theta \epsilon(\sigma),\qquad \delta\theta=\epsilon(\sigma),
\end{equation}
for some Grassmann $\epsilon(\sigma)$. We hope that this connection and the above mentioned symmetry might have an interesting new interpretation in the light of recent developments in WL/SA duality. 

\section{Supersymmetric Douglas functional and RNS amplitudes} \label{SDg}
We finish by demonstrating that exponent of the area of the cusped superworldsheet \eqref{SuperTdual} correctly reproduces RNS amplitudes. In order to evaluate the area we generalize  Douglas functional which allows to extract it directly from the boundary contour (in principle without the need of knowing the explicit solution). 

The algorithm for obtaining the functional is explained in detail in Appendix \ref{aA}. If we carefully repeat all the steps starting from \eqref{superA}, the resulting super-Douglas functional is given by
\begin{equation}
D[C]=\int dS_{1}\int dS_{2}\,C'^{\mu}(S_{1})C'^{\nu}(S_{2})\eta_{\mu\nu}\log\left(\Sigma(S_{1})-\Sigma(S_{2})\right),\label{SDouglas}
\end{equation}
where again $S_{i}=(\sigma_{i},\theta_{i})$ are the boundary points on the superworldsheet, boundary contour is denoted by $C^{\mu}(S)$, primes stand for $C'(S_i)=\partial_{\theta_{i}}C(S_i)+\theta_{i}\partial_{\sigma_{i}}C(S_i)$, and finally $\Sigma$ is the inverse of the reparametrization function of the boundary. Similarly to the bosonic case we supplement the functional with minimization equations that are obtained by varying \eqref{SDouglas} with respect to $\Sigma$.\\
To extract the required area we evaluate \eqref{SDouglas} on the supercontour 
\begin{equation}
C^{\mu}_{n}(S)=-\pi i\sum^{n}_{i=1}k^{\mu}_{i}\,\Theta(S-S_{i}).
\end{equation}
Again, in analogy with the bosonic case, derivatives of $C^{\mu}$ produce the supersymmetric delta functions. It is then easy to see that exponent of the Douglas functional correctly reproduces the integrand of the RNS amplitudes \cite{GS}
\begin{equation}
e^{\frac{D[C_{n}]}{4\pi^{2}}}=\prod^{n}_{i<j}(\sigma_{i}-\sigma_{j}-\theta_{i}\theta_{j})^{k_{i}\cdot k_{j}}.
\end{equation}
This is of course a good consistency check for the T-dual worldsheet $Y^{\mu}_{s}(\tau,\sigma,\theta)$ in \eqref{SuperTdual}. Again after fixing three of the $\sigma_{i}$ and inserting Mandelstam variables one could minimize this expression to obtain the superamplitude at high energies and fixed angle. This is equivalent to Douglas minimization.

\section{Remarks on super-Wilson loops}\label{FWL}
In this last section we make a few remarks/speculations on possible implications of our analysis on the relation between Wilson loops and scattering amplitudes. More precisely, we first argue that lightlike supercontours, when embedded in WL/SA story, naturally lead to a particular class of super-Wilson loops. Then, we briefly remind the reader of the definition and potential applications of such operators.
 
As mentioned in the introduction, in $\mathcal{N}$=4 SYM, lightlike Wilson loops are related to gluon scattering amplitudes\footnote{Since we try to keep the discussion as simple as possible, for all the omitted details and subtleties the reader is advised to consult e.g.\cite{Alday:2008yw,Alday:2010kn} and further references therein.}. The key object in this strong, as well as weak, coupling duality is the lightlike loop built of the physical momenta. Now, in this work, we have demonstrated that such loops can be easily generalized to superloops containing all the relevant information about semiclassical limit of the RNS amplitudes. With this information at hand, one obvious question to ask is: What would happen if we considered the $AdS$ setup of Alday and Maldacena with superstrings?

Following the ``historic path", we would start with applying the T-duality trick to the superstring disc amplitude. Based on the bosonic example, it is clear that we would then have to find a minimal supersurface attached to \eqref{Scontour} at the boundary of $AdS$. If that step was successful, and also the exponent of the minimal area correctly reproduced the strong coupling behavior of an appropriate (super) generalization of the gluon amplitude, we could claim that holographic super-Wilson loops are dual to the ``superamplitudes"\footnote{A precise meaning of this term in this context remains to be determined.}.

After that, our next natural step would be to check if the newly uncovered duality also holds in perturbation theory. In other words, we would embark on computing the expectation value of a Wilson loop along the lightlike supercontour \eqref{Scontour}.  This would bring us directly to a particular class of super-Wilson loops developed in a work of Andreev and Tseytlin \cite{Andreev:1988cb} (see also \cite{Migdal:1996rn} for extension to the momentum space); namely Wilson loops along supercontours with bosonic and fermionic paths. 

Now, even though most of the technical details behind the above steps are beyond the scope of this work, below we give a brief summary of the super-Wilson loop operators that deserve the attention on their own.

The super-Wilson loops (SWL) of our interest depend on both, bosonic $x^{\mu}(s)$ and fermionic $\psi^{\mu}(s)$ paths that can be elegantly packaged into one supercontour $X^{\mu}(s,\theta)=x^{\mu}(s)+\theta\,\psi^{\mu}(s)$, with Grassmann $\theta$. They can be defined in a complete analogy with the ordinary bosonic WL. Namely, while bosonic operators are path ordered $\mathcal{P}$ exponentials of a gauge field $A_{\mu}$ transported along $x^{\mu}(s)$
\begin{equation}
W[x]=\mathcal{P}\exp\left(i\int ds\,\dot{x}^{\mu}(s)A_{\mu}(x(s))\right),\label{BWL}
\end{equation}
SWL are given by
\begin{equation}
W_{s}[X]=\mathcal{P}_{s}\exp\left(i\int dS\,DX^{\mu}(S)A^{s}_{\mu}(X(S))\right),\label{SUPERWL}
\end{equation}
where, $S\equiv(s,\theta)$, $dS\equiv ds\,d\theta$ and $D=\partial_{\theta}+\theta\partial_{s}$. Now $A^{s}(X)$ is a superconnection and super-path ordering $\mathcal{P}_s$ is defined analogously to the bosonic counterpart (see e.g. \cite{Schubert:2001he}).

In order to understand \eqref{SUPERWL} from a more standard perspective, we can perform the Grassmann integral over $\theta$. Then the gauge invariant loop is given by
\begin{equation}
W[x,\psi]=\mathcal{P}\exp\left[i\int ds\,\left( \dot{x}^{\mu}(s)A_{\mu}(x(s))-\frac{1}{2}\psi^{\mu}(s)\psi^{\nu}(s)F_{\mu\nu}(x(s))\right)\right],\label{WLWL}
\end{equation}
where $A(x(s))$ is the bosonic connection and its field strength 
\begin{equation}
F_{\mu\nu}=\partial_{\mu}A_{\nu}-\partial_{\nu}A_{\mu}+[A_{\mu},A_{\nu}].
\end{equation} 
These SWLs are invariant under local supersymmetry (worldline SUSY) transformations
\begin{equation}
 \delta x^{\mu}=\psi^{\mu}(s)\epsilon,\qquad \delta\psi^{\mu}=\dot{x}^{\mu}(s)\epsilon
\end{equation}
for some constant Grassmann $\epsilon$.

Moreover, there also exists a very interesting geometrical connection between SWL and \eqref{BWL}. Namely they are related by the linear operator \cite{Migdal:1996rn}
\begin{equation}
W[x,\psi]=\mathcal{P}\exp\left(-\frac{i}{2}\int ds\,\psi^{\mu}(s)
\psi^{\nu}(s)\frac{\partial}{\partial \sigma_{\mu\nu}}\right)W[x],\label{Surface}
\end{equation}
where $\partial/\partial\sigma_{\mu\nu}$ is the area derivative of a loop; a well known object from the Makeenko-Migdal loop equations \cite{Makeenko:1979pb}.

Clearly, understanding the role of \eqref{WLWL} in WL/SA duality seems to be a promising direction. Not only establishing its precise amplitude dual but also finding the amplitude counterpart (at weak and strong coupling) of \eqref{Surface} might give a new perspective on this fascinating duality.

Last but not least, it is worth noting that \eqref{WLWL} usually appears as a part of worldline Lagrangians and plays an important role in string-inspired approaches to quantum field theory (see \cite{Schubert:2001he} for review on the subject). This link suggests that WL/SA might be thought of as an incarnation of such approaches.

\section{Conclusions and Comments}

In this Letter we investigated flat space, high energy scattering of open superstrings in RNS formulation. We demonstrated that, similarly to bosonic strings (and strings in AdS), a surface that captures all the relevant information about the process is spanned by the lightlike cusped supercontour for which the difference between neighboring cusps is proportional to the on-shell momentum.

We constructed a superworldsheet with bosonic and fermionic parts that separately satisfy the flat space T-duality constraints (Cauchy-Riemann equations). Then we verified that exponent of the area of such surface correctly reproduces the RNS amplitudes. It would be interesting to repeat (justify) the T-duality analysis of the classical superworldsheet in the Green-Schwarz formalism in the spirit of \cite{Berkovits:2008ic}. 

Our key tool in obtaining the area of the minimal surfaces was the flat space Douglas functional over reparametrizations of the boundary.  It computes the area of a surface directly from the boundary contour. An intuitive derivation of this functional is contained in Appendix \ref{aA}.  Clearly, if generalized to surfaces in curved backgrounds, such functional would be of a great importance for AdS/CFT (see \cite{Ambjorn:2011wz} for some recent progress on Douglas functional in $AdS$ spacetime). 

Finally, the main result of this Letter is the lightlike contour with fermions. This object,  parametrized in terms of the supersymmetric step function, appeared before in the context of super-Wilson loops in QCD \cite{Migdal:1996rn}. This link brought us to a class of super-Wilson loops that are not only the simplest supersymmetric extensions of the ordinary bosonic operators but are also related to the former via the area derivative commonly used in loop equations. As far as we are aware, operators of this precise form have not yet been explored in the context of WL/SA duality (except of a related discussion on worldline formalism in \cite{McGreevy:2008zy}). On the other hand, these SWL seem related to Wilson loops that contain the information about polarizations \cite{Mason:2010yk,CaronHuot:2010ek,Beisert:2012gb}. Most likely the operators mentioned here are just ``toy examples" for how extra structures might be included into the geometry of the lightlike loops. Nevertheless, a quantitative understanding of their role in $\mathcal{N}$=4 SYM might provide us with a broader perspective on the relation between Wilson loops and scattering amplitudes.

\vspace{1cm} \centerline{\bf Acknowledgments}

\vspace{0.5cm}
\noindent
We are grateful to Rutger Boels, Dimitrios Giataganas, Vishnu Jejjala, Yuri Makeenko, Thomas S{\o}ndergaard for many interesting discussions and  Robert de Mello Koch, Costas Zoubos and especially Shinji Hirano for reading the draft and very useful comments. This work is based upon research supported by the South African Research Chairs Initiative of the Department of Science and Technology and National Research Foundation.

\appendix

\section{Douglas functional}\label{aA}
Douglas functional is a well known tool in the field of the minimal surfaces. It was first introduced by Jesse Douglas in 1931 \cite{Douglas} as a key object in his solution to the Plateau problem which is to prove the existence of the surface with minimal area for a given boundary contour.  
In physics, Douglas approach to minimal surfaces was first advertised by Migdal in \cite{Migdal:1993sx} and more recently,it was employed in the studies of Regge amplitudes in QCD \cite{Makeenko:2008sh}. Derivation of the functional for bosonic surfaces was reviewed in \cite{Makeenko:2010dq} and in this appendix we derive the Douglas functional for supersurfaces. 

We start with the RNS superstring action in the superfield formulation 
\begin{equation}
S=\int\,d^{2}z d^{2}\theta \,\bar{D}X^{\mu}_{s}DX^{\nu}_{s}\eta_{\mu\nu}.
\end{equation}
We are interested in a minimal supersurface that solves the Dirichlet problem 
\begin{equation}
\bar{D}DX^{\mu}_{s}(y,x,\theta)=0,\qquad X^{\mu}_{s}(0,x,\theta)=C^{\mu}(x,\theta).
\end{equation}
Now we integrate by parts\footnote{Using $\int d\theta \partial_{\theta}V=0$ for every $V$.}
\begin{equation}
S=\int dT\,\left(X^{\mu}DX_{\mu}\right)_{y=0}.\label{SBDR}
\end{equation}
where for convenience we denote
\begin{equation}
\int dT=\int d\theta\int^{+\infty}_{-\infty}\,dx.
\end{equation}
Then we use the Poisson kernel on super-upper half plane 
\begin{equation}
P_{y}(S-S')=\frac{1}{\pi}\partial_{y}\log\sqrt{(S-S')^{2}+y^{2}}=\frac{1}{\pi}\frac{y}{(S-S')^{2}+y^{2}},
\end{equation}
where $S-S'$ the generalized distance between points at the superboundary
\begin{equation}
S-S'=x-x'-\theta\,\theta'.
\end{equation} 
A general solution the Dirichlet problem is given by
\begin{equation}
X^{\mu}_{s}(y,S)=\frac{1}{\pi}\int dT\,\frac{C^{\mu}(T)\,y}{(S-T)^{2}+y^{2}}.
\end{equation}
Finally inserting it to \eqref{SBDR}, allowing for an arbitrary reparametrization function $F(T)$, and changing the (super)integration variables we arrive at
\begin{equation}
D_{s}[C]=\int dF_{1}\int dF_{2}\,C'^{\mu}(F_{1})C'^{\nu}(F_2)\,\eta_{\mu\nu}\log\left(\Sigma(F_1)-\Sigma(F_2)\right).\label{APD}
\end{equation}
If we vary with respect to $\Sigma$ we get the super minimization conditions
\begin{equation}
\int dF_{1}\,\frac{C'^{\mu}(F_{1})C'_{\mu}(F_{2})}{\Sigma(F_{1})-\Sigma(F_{2})}=0.
\end{equation}
Naturally in order to keep track of bosonic and fermionic contributions at each step separately,  one can just expand all the formulas to the leading order in $\theta$s. For instance, inserting 
\begin{equation}
C^{\mu}(F_i)\equiv x^{\mu}(f_i)+\theta_i\,\psi^{\mu}(f_i),
\end{equation}
into \eqref{APD}, recalling that $C'(F_i)=(\partial_{\theta_i}+\theta_i\partial_{f_i}) C(F_i)$, and finally performing the Grassmann integration, leaves us with the component form of the super-Douglas functional
\begin{equation}
D[x,\psi]=\int df_{1}\int df_{2}\left(x'(f_{1})\cdot x'(f_2)\log\left(\sigma(f_1)-\sigma(f_2)\right)-\frac{\psi(f_1)\cdot \psi(f_2)}{\sigma(f_1)-\sigma(f_2)}\right).
\end{equation}

\end{document}